# Large Enhancement of the Photoluminescence Emission of Photoexcited Undoped GaAs Quantum Wells Induced by an Intense Single-Cycle Terahertz Pulse

K. Shinokita[1], H. Hirori[2,*], and K. Tanaka[1,2]

[1] *Department of Physics, Graduate School of Science, Kyoto University,*

*Kyoto 606-8502, Japan*

[2] *Institute for Integrated Cell-Material Sciences (WPI-iCeMS), Kyoto University, and*

*JST-CREST, Kyoto 606-8501, Japan*

T. Mochizuki, C. Kim, and H. Akiyama

*Institute for Solid State Physics, University of Tokyo, and JST-CREST,*

*Chiba 277-8581, Japan*

L. N. Pfeiffer and K. W. West

*Department of Electrical Engineering, Princeton University, Princeton, New Jersey*

*08544, USA*

## Abstract

Intense terahetz (THz) pulses induce a photoluminescence (PL) flash from undoped high-quality GaAs/AlGaAs quantum wells under continuous wave laser excitation. The number of excitons increases 10000-fold from that of the steady state under only laser excitation. The THz electric field dependence and the relaxation dynamics of the PL flash intensity suggest that the strong electric field of the THz pulse ionizes impurity states during the one-picosecond period of the THz pulse and release carriers from a giant reservoir containing impurity states in the AlGaAs layers.

The topic of photoexcited carrier dynamics in quantum wells (QWs) has attracted considerable attention from researchers interested in technological applications and fundamental physics [1-3]. Owing to their direct bandgaps and high electron mobilities, compound semiconductor QWs, e.g., GaAs and AlGaAs, have advantages for applications ranging from efficient photovoltaic devices to radio-frequency electronics and optoelectronics [3]. Time-resolved photoluminescence (PL) measurements have been extensively used to study the lifetimes and relaxation mechanisms of carriers in QWs [1,4-6]. In these measurements, the relaxation process depends on the initial distribution of the electron-hole pairs created by the photoexcitations or injected current [6,7], and the carriers created in the barrier or cap layers, i.e., non-active layers, become trapped at various local sites originating from chemical, structural, and interface imperfections. Although these trapped carriers have a significant influence on the carrier dynamics and performance of devices [8-10], their origin and distribution are as yet unknown. Recently, it has been demonstrated that a tunable, narrowband terahertz (THz) wave source induces a photoionization of trap states and is useful for studying them, because their specific excitation energies may lie in the few meV or THz spectral region [11-13]. Moreover, the ultra-intense electric fields of single-cycle THz pulses [14-17] have been shown to induce field-ionization processes in trap states [18,19]. Thus, these phenomena may give us complementary information on the fundamental characteristics of trap states, which cannot be obtained with optical methods, and enable us to study their influence on nonlinear transport and phase transition phenomena under instantaneous high-electric fields [18-21].

We studied photoexcited carrier dynamics in undoped high-quality GaAs/AlGaAs quantum wells by using time-resolved photoluminescence spectroscopy involving the simultaneous use of a continuous-wave visible laser and intense single-cycle THz pulse excitations. The THz pulses induce bright PL flashes of exciton emissions from GaAs wells under visible continuous wave (CW) laser excitation. The THz electric field dependence and relaxation dynamics of the PL flash intensity suggest that the strong electric field of the THz pulse ionizes impurity states and releases carriers from a giant reservoir containing many impurity states in the AlGaAs layers, which cannot be observed by optical methods.

Figure 1(a) shows a schematic diagram of the experimental setup. The THz pulses were generated by optical rectification of femtosecond laser pulses in LiNbO$_3$ crystal by using the tilted-pump-pulse-front scheme [15,17]. The spectrum of the THz pulse has a maximum intensity around 0.8 THz and a bandwidth of ~1.5 THz (full width at half-maximum intensity (FWHM)) [17]. An amplified Ti:sapphire laser (repetition rate: 1 kHz, central wavelength: 780 nm, pulse duration: 100 fs, and 4 mJ/pulse) was used as the laser pulse source. A pair of wire-grid polarizers was used to vary the field amplitude of the THz pulse without modifying its waveform.

The multiple quantum well (QWs) sample consisted of a stack of ten 10-nm-wide undoped GaAs wells separated by 12-nm-wide Al$_{0.3}$Ga$_{0.7}$As barriers grown via molecular-beam epitaxy (MBE) on a (001)-GaAs substrate (thickness of 370 μm), and the QWs were sandwiched between 1-μm Al$_{0.33}$Ga$_{0.67}$As digital-alloy-barrier layers. The high quality of the sample with a low impurity concentration of <10$^{15}$ cm$^{-3}$ was attested by mobilities of ~2×10$^6$ cm$^2$V$^{-1}$s$^{-1}$ measured on other heterostructures grown in the same MBE system shortly thereafter. The sample was mounted in vacuum on a liquid-helium-cooled cold finger, cooled down to 10 K, and continuously excited by the CW laser (2.33 eV) above the bandgap energy of the AlGaAs layers (1.88 eV). Along with the visible laser excitation, an intense THz pulse with a maximum peak-electric-field amplitude of 0.77 MV/cm irradiated the sample. The visible laser beam was directed through a small hole in the parabolic mirror used to focus the THz pulse. The spot sizes of the visible laser and THz pulse on the sample were ~100 and ~300 μm in diameter, respectively. A Hamamatsu streak camera coupled to a spectrometer with a resolution of about ~1 meV was used to make time- and wavelength-resolved detections of the PL induced by the THz pulse.

Figures 2(a) and 2(b) show a typical streak camera image and its spectrally integrated transient of the PL from the QWs excited by the visible CW laser and THz pulse. The steady-state PL intensity $I_{PL}$ measured only with the visible laser excitation (0.6 W/cm$^2$) is very weak (blue solid line in Fig. 2(b)). Nonetheless, a bright PL flash arises just after the arrival of the THz pulse. Figure 2(c) shows the time integrated spectra of the PL flash shown in Fig. 2 (a) and the PL measured with only the visible CW laser excitation. The peak energies centered around 1.54 eV are almost the same, and Kronig–Penney analysis shows that their origins can be attributed to $n$=1 heavy-hole exciton emissions in the wells [22].

Figure 3(a) shows the CW laser power dependence of the PL flash intensity induced by a 0.77-MV/cm THz pulse. With increasing visible CW laser intensity, the flash intensity increases and shows a saturation behavior. In the minimum photoexcitation intensity case (2.33 eV, 60 μW/cm$^2$) without THz pulse irradiation, the generated exciton density $N_{ex}$ in the steady state and the total exciton density $N_{total}$ during the 1-ms repetition period $T_{rep}$ in the GaAs wells are estimated to be 2.5×10$^3$ cm$^{-2}$ and 1.3×10$^9$ cm$^{-2}$ (=$N_{ex}T_{rep}/\tau_{ex}$, where $\tau_{ex}$ is exciton lifetime) [23]. The PL flash amplitude reaches 1.2×10$^4$-fold that of the steady-state PL intensity, and the increased exciton density $\Delta N$ in the PL flash is estimated to be ~3×10$^7$ cm$^{-2}$. This means that the THz pulse irradiation can transiently increase the number of excitons by four orders of magnitude with respect to the steady-state exciton density $N_{ex}$, and the PL yield increment of the GaAs wells is estimated to be ~10$^{-2}$ (=$\Delta N/N_{total}$).

Figure 3(b) shows the THz pump electric field dependence of the flash PL intensity. As the electric field increases, the PL flash intensity increases gradually and starts to saturate at around 0.5 MV/cm. The released carrier density can be theoretically estimated by assuming a field-assisted tunneling process as a way of trap ionization. Additionally, we found that the PL flash is absent for the excitations (1.58 eV) below the bandgap of AlGaAs layers having various excitation densities, indicating that the trap states exist in the AlGaAs layers. The time evolution of the trapped carrier density $N_t(t)$ under THz pulse irradiation can be described by [24]:

$$\frac{dN_t(t)}{dt} = -\tau^{-1} N_t(t), \tag{1}$$

$$\tau^{-1}(E_{THz}^{in}(t)) = \frac{eE_{THz}^{in}(t)}{2\sqrt{2m^* I_B}} \exp\left(-\frac{4}{3} \frac{I_B^{\frac{3}{2}} \sqrt{2m^*}}{\hbar e E_{THz}^{in}(t)}\right),$$

where $\tau$ is the tunneling time of carriers, and $e$ is the charge of the electron, and $I_b$ and $m^*$ are the binding energy of trap states and effective mass of the electron (0.072 $m_0$) or light hole (0.105 $m_0$) in the AlGaAs layers [25]. $E_{THz}^{in}(t)$ is the temporal THz electric field inside the sample [18]. By integrating Eq. (1), the density of remaining trapped carriers after the THz pulse excitation can be represented as follows:

$$N_t = N_0 - \int_{-\infty}^{\infty} \tau^{-1}(t) N_t(t) dt. \tag{2}$$

Here, the conservation condition, $N_0 = N_t + N_r$, is assumed, where $N_0$ and $N_r$ are respectively the total trap state density and released carrier density. Using the temporal profile of the THz electric field $E_{THz}(t)$, we can numerically calculate the released carrier density $N_r$ from Eq. (2).

Assuming the PL flash intensity $I$ is proportional to the released carrier density $N_r$, fitting Eq. (2) with a light-hole mass of 0.105 $m_0$ to the data shown in Fig. 3(b) indicates that the THz field dependence of the PL flash intensity is well described by a tunneling process and the best fit to the experimental data is obtained with $I_b$=40 meV [26]. This deduced binding energy is similar to that of shallow acceptors (~30–50 meV) of group IV elements like carbon and silicon substitute for arsenic atoms [25, 28], whereas that deduced from the electron mass ($I_b$=46 meV) does not match the donor binding energy (~10 meV) [25,29]. This result implies that the photoexcited carriers can be trapped at such shallow impurity donors and acceptors, and the field dependence of PL flash shown in Fig. 3(b) reflects a field ionization of accepter because of its larger binding energy. As shown in the inset of Fig. 3(b), the temperature dependence of the PL flash is similar to that of the PL intensity obtained from only the 2.33 eV-cw-laser excitation. This indicates that the gradual temperature dependence of the PL flash is governed by thermalization of excitons to large *k* states, which suppresses radiative recombination of excitons in GaAs QWs [5], and not by thermal ionization of the impurity states because of their larger binding energy $I_b$ compared with the thermal energy of this temperature regime.

The dynamics of the trapped carriers were studied by analyzing the temporal response of the PL flash as a function of the delay between the THz and visible-laser pulses shown in Fig. 4. In this experiment, the visible CW laser beam (2.33 eV, 60 W/cm$^2$) was modulated with an acousto-optic modulator (pulse width of 10 ns and repetition rate of 1 kHz), which was operated synchronously with the 1-kHz amplified Ti: sapphire laser clock and THz pulse source. Here, the visible laser pulse and THz electric field were held at 1 μJ/cm$^2$ and 0.77 MV/cm, respectively. The delay time between the visible and THz pulses was varied over a nanosecond to submillisecond time region by using an electronic-delay-generator device. Figure 4 shows the PL flash intensity as a function of the delay time, which is proportional to the number of carriers released by the THz pulse [27]. The intense THz pulses induce the PL flash after the visible laser pulse excitation, and the flash intensity decreases as the delay increases. The decay times of the fast and slow components are estimated to be $\tau_f \sim 0.3$ μs and $\tau_s \sim 1.6$ ms by using a bi-exponential function, i.e., $A_f \exp(-t/\tau_f) + A_s \exp(-t/\tau_s)$. Here, $A_f$ is 0.45, and and $A_s$ is 0.43. The long decay time of the PL flash suggests that the lifetime of photoexcited electron-hole pairs trapped at the shallow impurity states became longer as their spatial separation increased [30], much like the case of donor-acceptor (DA) pairs.

In contrast to the case of the PL flash, the excitation-power dependence of the DA PL intensity obtained from photoexcitation measurements without THz pulse excitation does not show any saturation behavior [31]. The saturation effect might occur if the impurity states available to bind carriers are limited; in fact, the estimated total impurity density of $\sim 10^9$ cm$^{-2}$ is much smaller than the total incident photon density ($\sim 10^{11}$-$10^{17}$ cm$^{-2}$) of the 2.33-eV excitation laser during the one-millisecond time period for the whole range from the minimum to maximum excitation intensity (60 μW/cm$^2$-60 W/cm$^2$) [32]. Additionally, the decay time of the impurity states involved in the PL flash is longer than one millisecond as shown in Fig. 4, which is particularly suitable for the saturation behavior. From these considerations, we conclude that during the 1-ms repetition period of the THz pulses, carriers excited by the CW laser accumulate in impurity states, and the intense THz pulse ionizes them in one burst [33]. The released carriers diffuse to the GaAs well layers and therein form excitons which recombine in a radiative process, leading to the bright PL flash, as shown in Fig. 2.

In summary, we studied the photoexcited carriers of undoped GaAs/AlGaAs QWs by using time-resolved photoluminescence spectroscopy with simultaneous visible laser and intense single-cycle THz pulse excitations. The bright PL flashes from the GaAs wells indicate that the THz pulse increases the number of excitons due to the field-ionization of the impurity states by up to 10000-fold that of the steady state. Our results shed light on new aspects of photoexcited carrier dynamics in high-quality QWs and their influence on nonlinear transport phenomena under instantaneous THz high-electric fields. As such, they might lead to improved electronic and optoelectronic devices. In addition, they suggest that the efficient erasing of trapped carriers during the one-picosecond period rise to a delayed PL flash that can be used for storage and retrieval of light [34].

**Acknowledgements**

We thank Takao Aoki for letting us use the optical pulse source with an acoustic modulator. H.H. thanks Alexej Pashkin, Alfred Leitenstorfer, Stephan W. Koch, Mackillo Kira for invaluable discussions during his visit in Konstanz through the iCeMS-JSPS Overseas Visit Program for Young Researchers (Bon Voyage Program), and Yoshihiko Kanemitsu for stimulating discussions. This study was supported by KAKENHI (24760042 and 20104007) from JSPS and MEXT of Japan, and Industry-Academia Collaborative R&D from Japan Science and Technology Agency (JST). This study was partly supported by KAKENHI (20104004 and 23360135) from MEXT and JSPS, and also by the Gordon and Betty Moore Foundation and the NSF MRSEC program through the Princeton Center for Complex Materials (DMR-0819860).

**Figure Captions**

**Fig. 1 (Color online)** Schematic experimental setup for photoluminescence flash induced by intense THz pulse. Generated THz pulses are focused onto the GaAs QW sample, and the luminescence is detected by a Hamamatsu streak camera coupled to a spectrometer with a resolution of about 1 meV. The visible laser beam was directed through a small hole in the parabolic mirror used to focus the THz pulse.

**Fig. 2 (Color online)** (a) Typical streak camera image of the photoluminescence flash from the QWs excited at 10 K by both the visible CW laser and THz pulse. The visible laser intensity is 0.6 W/cm$^2$ and the peak electric field of THz pulse is 0.77 MV/cm. (b) Spectrally integrated transients of the photoluminescence flash with the THz pulse in (a) (red solid line) and without the THz pulse (blue dashed line). The dashed arrow indicates the arrival time of the THz pulse. (c) Spectra of the time-integrated PL flash shown in (a) (red solid line) and the PL emission excited by only the visible CW laser with an intensity of 0.6 W/cm$^2$ (blue dashed line).

**Fig. 3 (Color online)** (a) Visible CW laser power dependence of the flash intensity induced by a 0.77-MV/cm THz pulse (solid circle). The solid line to guide the eye represents the 0.2-power-law intensity dependence. (b) THz electric field dependence of the flash PL intensity (solid circle). The solid curve is a theoretical calculation assuming a field-assisted direct-tunneling process described by Eqs. (1) and (2). The inset shows temperature dependence of PL excited by a 2.33 eV-CW-laser and PL flash.

**Fig. 4 (Color online)** Photoluminescence flash intensity induced by an intense THz pulse as a function of delay time between the visible and THz pulses. The experimental conditions are described in the body of the paper. The solid line is the fitting curve of a bi-exponential function.

**Fig.1**

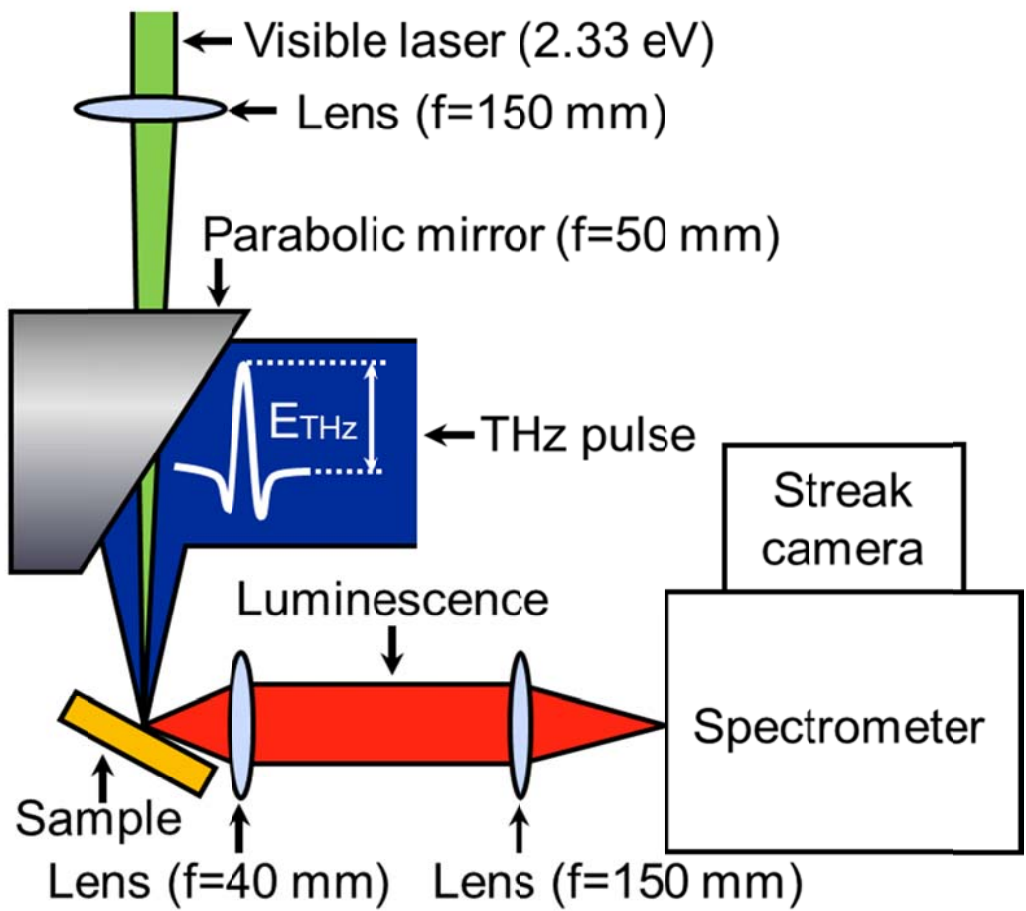

**K. Shinokita**

**Fig. 2**

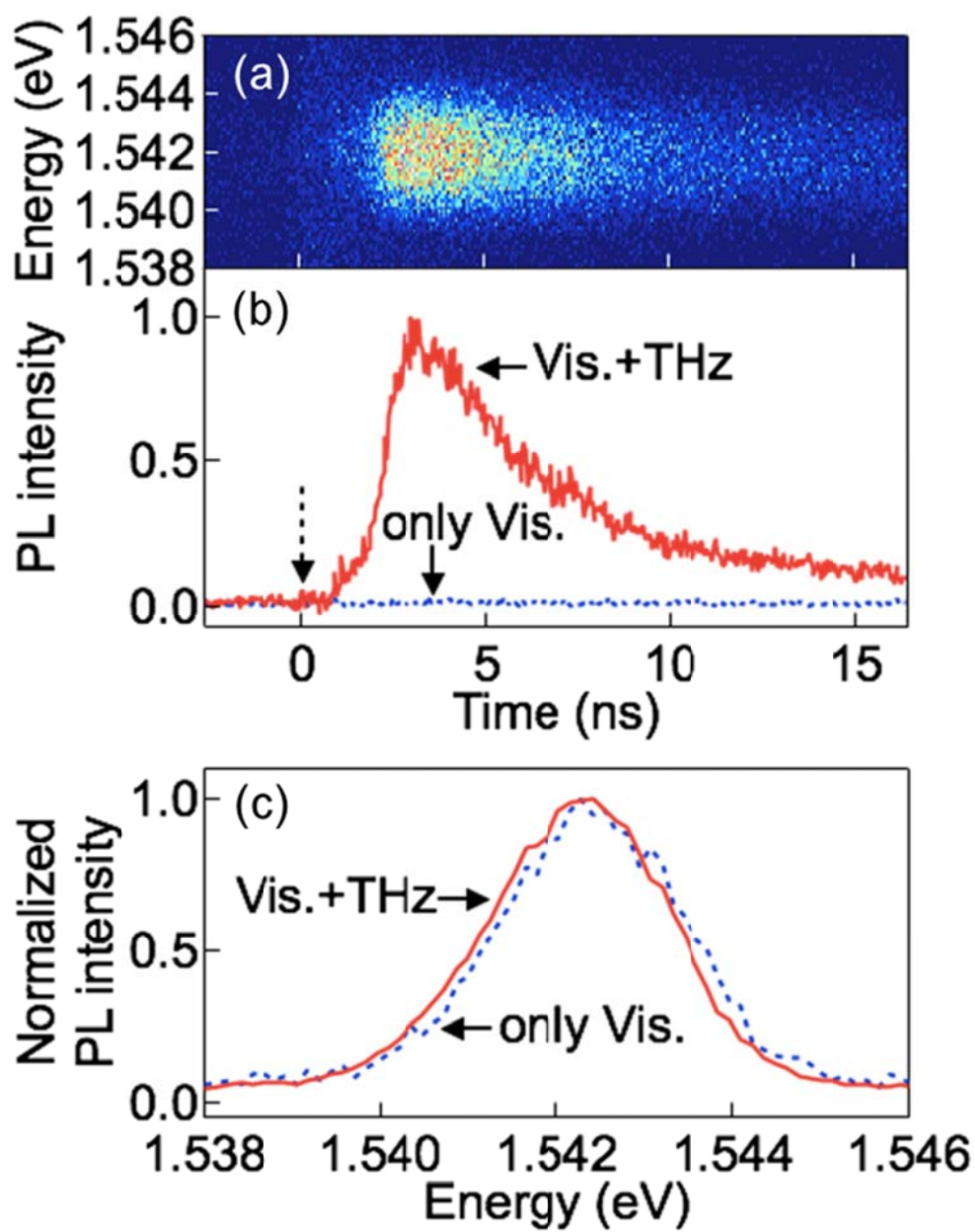

K. Shinokita

**Fig. 3**

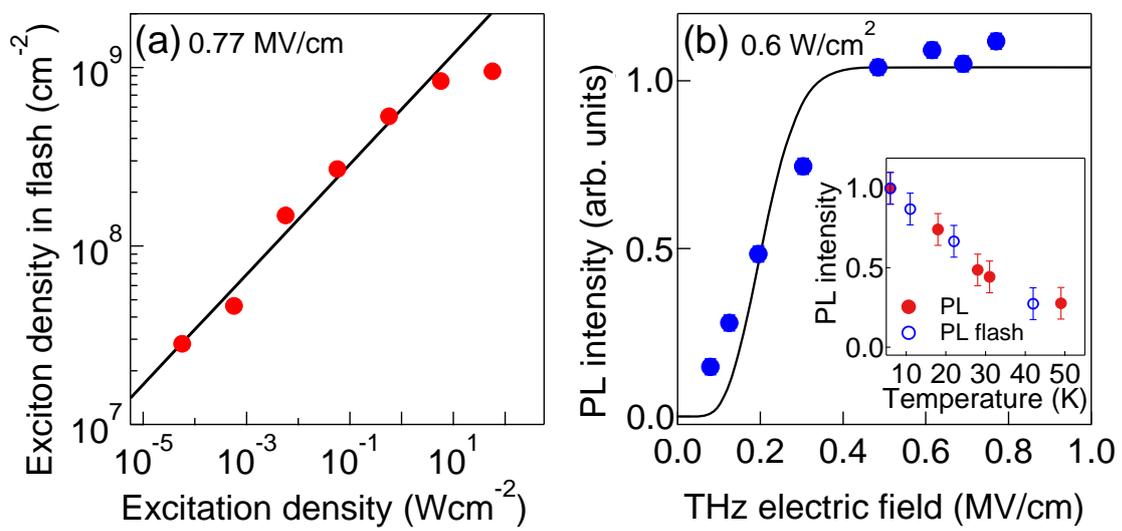

**K. Shinokita**

**Fig. 4**

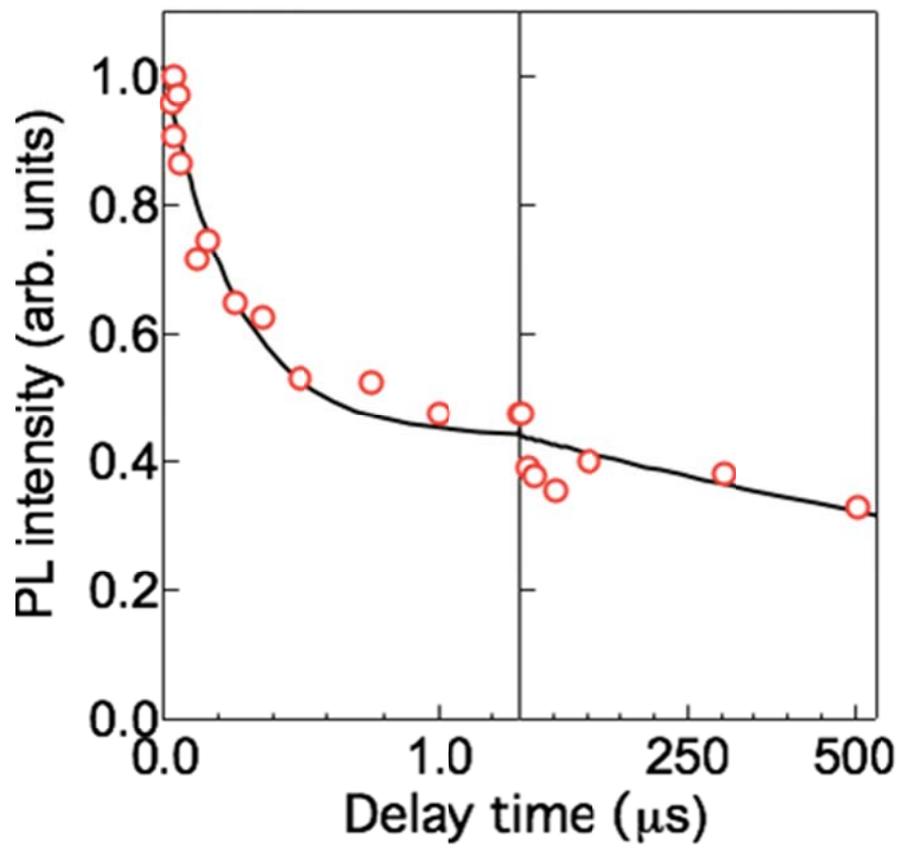

**K. Shinokita**